\documentstyle[11pt]{article}

\begin{document} 
\centerline{\bf Simulation of an entangled state in a chain of three nuclear spins system  }
\vskip2pc
\centerline{G.V. L\'opez and L. Lara}
\centerline{Departamento de F\'{\i}sica, Universidad de Guadalajara}
\centerline{Apartado Postal  4-137, 44410 Guadalajara, Jalisco, M\'exico}
\vskip2pc
\centerline{PACS: 03.67.Lx, 03.65.Ta}
\vskip2cm
\centerline{ABSTRACT}
\vskip1pc\noindent
We study the formation of an entangled state in a one-dimensional chain of three nuclear spins system
which interact weakly through the Ising type of interaction and taking into account first and second
neighbor  interactions. We can get this entangled state using two pulses ($\pi/2$ and $\pi$ pulses), and
we study the efficiency of getting this entangled state as a function of the ratio of the second neighbor
interaction coupling constant to the first neighbor interaction coupling constant ($J'/J$). We found that
for $J'/J\ge 0.04$, the entangled state is well defined.
\vfil\eject\noindent 
{\bf 1. Introduction}
\vskip0.5pc\noindent
The huge interest in quantum computation and quantum information was triggered by the polynomial time
solution of the prime factorization problem (Shor's algorithm [1]), the fast data base searching
(Grover's algorithm [2]), error correction codes [3], robust entanglement [4], and the teleportation
phenomenum [5]. Almost any quantum system with at least two quantum levels may be used, in principle,
for quantum computation. This one uses qubits (quantum bits) instead of bits to process information. A
qubit is the superposition of any two levels of the system , called $|0\rangle$ and $|1\rangle$ states,
$\Psi=C_0|0\rangle+C_1|1\rangle$ with $|C_0|^2+|C_1|^2=1$. The tensorial product of $L$-qubits makes up
a register of lenght $L$, say $|x\rangle=|i_{L-1},\dots,i_0\rangle$, with $i_j=0,1$, and a quantum
computer with $L$-qubits works in a $2^L$ dimensional Hilbert space, where an element of this space is
of the form $\Psi=\sum C_x|x\rangle$, with $\sum|C_x|^2=1$.  
\vskip0.5pc\noindent
Quantum computers of few qubits [6] have been in operations and have been used to explore quantum gates,
entanglement, small number Shor's factorization, small data base Grover's searching, teleportation, error
corrections, and cryptography. However, to make serious computer calculations one may requires a quantum
computer with at least of 100-qubits registers, and we think that will be hopefully achived in a near
future. One solid state quantum comuter model that has been explored for physical realization and which
allows to make analytical studies is that one made up by one-dimensional chain of nuclear spins sytems [7]
, where the Ising interaction among first neighbor spins allows to inplement  ideally this type of
computer  with 1000-qubits or more [8]. One of the important phenomena we studied with this model was the
entangled state formation [9]. In this paper, we consider second neighbor spin interaction in a chain of
three nuclear spins system. We show that this  allows us to implement an entangled state using two pulses
($\pi/2$ and $\pi$), and we determine the threshold of the second neighbor interaction coupling
parameter to get a well define entangled state.

\vskip2pc
\leftline{\bf 2. Equation of Motion}
\vskip1pc\noindent
Consider a one-dimensional chain of three equally spaced nuclear-spins system (spin one half) making an
angle $\cos\theta=1/\sqrt{3}$ with respect the z-component of the magnetic field (selected in this way to
eliminate the dipole-dipole interaction between spins) and having an rf-magnetic field in the transversal
plane. The magnetic field is given by
$${\bf B}=(b\cos(\omega t+\varphi), -b \sin(\omega t+\varphi), B_o(z))\ ,\eqno(1)$$
where $b$  is the amplitude of the rf-field, $B_o(z)$ is the amplitude of the z-component of the magnetic
field, $\omega$ and $\varphi$ are the angular frequency and phase of the rf-field. So, the Hamiltonian
of the system is given by
$$H=-\sum_{k=0}^2{\bf \mu_k}\cdot {\bf B_k}-2J\hbar\sum_{k=0}^1I_k^zI_{k+1}^z
-2J'\hbar\sum_{k=0}^0I_k^zI_{k+2}^z\ ,\eqno(2)$$
where ${\bf\mu_k}$ represents the magnetic moment of the kth-nucleus which is given in terms of the
nuclear spin as ${\bf\mu_k}=\hbar\gamma(I_k^x,I_k^y, I_k^z)$, being $\gamma$ the proton gyromagnetic
ratio. ${\bf B_k}$ represents the magnetic field at the location of the $kth$-spin ($z=z_k$). The second
term at the right side of (2)  represents the  first neighbor spin interaction, and the third term
represents the second neighbor spin interaction. $J$ and $J'$ are the coupling constants for these
interactions. This Hamiltonian can be written in the  following way
$$H=H_0+W\ ,\eqno(3a)$$
where $H_0$ and $W$ are given by
$$H_0=-\hbar\left\{\sum_{k=0}^2\omega_kI_k^z+2J(I_0^zI_1^z+I_1^zI_2^z)+2J'I_0^zI_2^z\right\}\eqno(3b)$$
and
$$W=-{\hbar\Omega\over 2}\sum_{k=0}^2\biggl[e^{i(\omega t+\varphi)}I_k^++e^{-i(\omega
t+\varphi)}I_k^-\biggr]\ ,\eqno(3c)$$ 
where $\omega_k=\gamma B_o(z_k)$ is the Larmore frequency of the
kth-spin, $\Omega=\gamma b$ is the Rabi's frequency, and $I_k^{\pm}=I_k^x\pm iI_k^y$ represents the ascend
operator (+) or the descend operator (-). The Hamiltonian $H_0$ is diagonal on the basis
$\{|i_2i_1i_0\rangle\}$, where $i_j=0,1$ (0 for the ground state and 1 for the exited state),
$$H_0|i_2i_1i_0\rangle=E_{i_2i_1i_0}|i_2i_1i_0\rangle\ .\eqno(4a)$$
The eigenvalues $E_{i_2i_1i_0}$ are given by
$$E_{i_2i_1i_0}=-{\hbar\over 2}\biggl\{(-1)^{i_2}\omega_2+(-)^{i_1}\omega_1+(-1)^{i_0}\omega_0+
J[(-1)^{i_0+i_1}+(-1)^{i_1+i_2}]+(-1)^{i_0+i_2}J'\biggr\}\ .\eqno(4b)$$
The term (3c) of the Hamiltonian (3a) allows to have a single spin transitions on the above eigenstates
by choosing the proper resonant frequency, as shown in Figure 1. In this work, we are  interested in
the transitions $|000\rangle\longleftrightarrow |001\rangle$ and $|001\rangle\longleftrightarrow
|101\rangle$ which have the resonant frequencies 
$$\omega=\omega_0+J+J'\eqno(5a)$$
and
$$\omega=\omega_2+J-J'\ .\eqno(5b)$$
To solve the Schr\"odinger equation
$$i\hbar{\partial\Psi\over\partial t}=H\Psi\ ,\eqno(6)$$ 
let us propose a solution of the form
$$\Psi(t)=\sum_{k=0}^7C_k(t)|k\rangle\ ,\eqno(6)$$
where we have used decimal notation for the eigenstates in (4a), $H_0|k\rangle=E_k|k\rangle$.
Substituting (6) in (5), multiplying for the bra $\langle m|$, and using the orthogonality relation 
$\langle m|k\rangle=\delta_{mk}$, we get the following equation for the coefficients
$$i\hbar\dot C_m=E_mC_m+\sum_{k=0}^7C_k\langle m|W|k\rangle\ \ m=0,\dots,7.\eqno(7)$$
Now, using the following transformation
$$C_m(t)=D_m(t)e^{-iE_m t/\hbar}\ ,\eqno(8)$$
the fast oscillation term $E_mC_m$ of Eq. (7) is removed (this is equivalent to going to the interaction
representation), and the following equation is gotten for the coefficients $D_m$
$$i\dot D_m={1\over\hbar}\sum_{k=0}^7W_{mk}D_ke^{i\omega_{mk}t}\ ,\eqno(9a)$$
where $W_{mk}$  denotes the matrix elements $\langle m|W|k\rangle$, and $\omega_{mk}$ are defined as
$$\omega_{mk}={E_m-E_k\over\hbar}\ .\eqno(9b)$$
Eq. (9a) represents a set  of sixteen real coupling ordinary differential equations which can be solved
numerically, and where $W_{mk}$ are the elements of the matrix
$$(W)=-{\hbar\Omega\over 2}\pmatrix{
0 & z^* & z^* & 0   & z^* & 0   & 0   & 0 \cr
z &  0  &  0  & z^* & 0   & z^* & 0   & 0 \cr
z &  0  &  0  & z^* & 0   & 0   & z^* & 0 \cr
0 &  z  &  z  &  0  & 0   & 0   & 0   & z^*\cr
z &  0  &  0  &  0  &  0  & z^* & z^* & 0  \cr
0 &  z  &  0  &  0  & z   &  0  & 0   & z^*\cr
0 &  0  &  z  &  0  &  z  &  0  & 0   & z^*\cr
0 &  0  &  0  &  z  &  0  &  z  & z   & 0  \cr}\ ,\eqno(9c)$$
where $z$ is defined as $z=e^{i(\omega t+\varphi)}$, and $z^*$ is its complex conjugated.
\vskip2pc
\leftline{\bf 3. Numerical Simulations}
\vskip1pc\noindent
We start with the ground state, $\Psi_0=|000\rangle$, of the system and apply a $\pi/2$-pulse with
$\varphi=0$ and with frequency $\omega=\omega_0-J-J'$ to get the superposition state
$$\Psi_1={1\over\sqrt{2}}\biggl(|000\rangle+|001\rangle\biggr)\ .\eqno(10)$$
Then, we apply a $\pi$-pulse with $\varphi=0$ and frequency $\omega=\omega_2+J-J'$ to get the entangled
state
$$\Psi_2={1\over\sqrt{2}}\biggl(|000\rangle-|101\rangle\biggr)\ .\eqno(11)$$
The entangled state with plus sign can be gotten using a phase $\varphi=\pi$.
To solve numerically (9a), we select similar values for the parameters as reference 8 and 9. So, in
units of $2\pi\times MHz$, we set the following values
$$\omega_0=100\ ,\ \omega_1=200\ ,\ \omega_2=400\ ,\ J=5\ ,\ \Omega=0.1\eqno(12)$$
The coupling constant $J'$ is chosen with at least one order of magnitude less that $J$ since in the
chain of spins one expect that second neighbor contribution to be at least one order of magnitude weaker
than first neighbor contribution, depending on the interspace separation of the nuclei. In all our
simulations the total probability,
$\sum|C_k(t)|^2$, is conserved equal to one within a precision of $10^{-6}$. 
\vskip0.5pc\noindent
Figure 2 shows the behavior of $Re~ D_0$, $Im~D_0$, $Re~D_5$, and $Im~D_5$  during the two pulses
($t=\tau=\pi/2\Omega+\pi/\Omega$) for the digital initial state and with $J'=0.2$. We can see the
formation of the superposition state after the first $\pi/2$-pulse and the formation of the entangled
state (11) after the following $\pi$-pulse. Fig. 3 shows the behavior of the probabilities $|C_k|^2$,
$k=0,\dots,7$ during the two pulses with the clear formation of the superposition state and the entangled
state. Fig. 4 shows the behavior of the expected z-component of the spin,
$$\langle I_0^z\rangle={1\over 2}\sum_{k=0}^7(-1)^k|C_k(t)|^2\ ,\eqno(13a)$$
$$\langle I_1^z\rangle={1\over
2}\biggl\{|C_0|^2+|C_1|^2-|C_2|^2-|C_3|^2+|C_4|^2+|C_5|^2-|C_6|^2-|C_7|^2\biggr\}\ ,\eqno(13b)$$
and
$$\langle I_2^z\rangle={1\over 2}\sum_{k=0}^3|C_k|^2-\sum_{k=4}^7|C_k|^2\ ,\eqno(13c)$$
during the two pulses. As one could expect $\langle I_0^z\rangle=\langle I_2^z\rangle=0$ at the end of the
two pulses due to the formation of the entangled state (11). The expected value of the spin is rotating
in the plane $(x,y)$ as is shown on Fig. 5. These transversal expected values are given by
$$\langle I_0^x\rangle=Re\biggl(C_1^*C_0+C_3^*C_2+C_5^*C_4+C^*_7C_6\biggr)\ ,\hskip1cm\langle
I_0^y\rangle=Im\biggl(\dots\biggr)\ ,\eqno(14a)$$
$$\langle I_1^x\rangle=Re\biggl(C_2^*C_0+C_3^*C_1+C_6^*C_4+C^*_7C_5\biggr)\ ,\hskip1cm\langle
I_1^y\rangle=Im\biggl(\dots\biggr)\ ,\eqno(14b)$$ 
and
$$\langle I_2^x\rangle=Re\biggl(C_4^*C_0+C_5^*C_1+C_6^*C_2+C^*_7C_3\biggr)\ ,\hskip1cm\langle
I_2^y\rangle=Im\biggl(\dots\biggr)\ .\eqno(14c)$$
To determine the range of values of
$J'$ (second neighbor coupling constant) for which the entangled state is well defined after the two
pulses, that is, where the other resonances and non-resonant transition are canceled, we calclate the
fidelity paramete [10] for this process,
$$F=\langle\Psi_{expected}|\Psi_{numerical}\rangle\ ,\eqno(15)$$
where $\Psi_{expected}$ is our state (11), and $\Psi_{numerical}$ is the resulting wave function from
our simulations. Fig. 6 shows the fidelity as a function of the ratio of the second neighbor interaction
constant to first neighbor interaction constant, $J'/J$. As one can see, for a value $J'/J\ge 0.04$ one
gets a very well defined entangled state. This means that the second neighbor interaction with even two
orders of magnitud weaker than the first neighbor interaction, we can generate an entangled state in this
system.
\vfil\eject
\vskip3pc\noindent
\leftline{\bf 4. Conclusions and Comments}
\vskip0.5pc\noindent
We have studied the formation of an entangled state using two pulses in a chain of three nuclear spins
system with first and second neighbor Ising spin interaction. The characteritics of the entangled state
were determinated, and we found that the entanglement can be realized even for very weak second neighbor
spin interaction ($J'/J\ge 0.04$). We consider that the coupling constant $J'$ may play an important
rolle in the so called $2\pi$-method found in reference [8,9] to supress non-resonant transition in the
chain of nuclear spins system  because this parameter will enter in the detuning parameter (
$\Delta=(E_p-E_m)/\hbar-\omega$).
\vskip5pc\noindent
\leftline{\bf Acknowledgements}
\vskip0.5pc\noindent
 This work was supported by SEP under the contract PROMEP/103.5/04/1911 and the University of Guadalajara.
\vfil\eject
\leftline{\bf Figure Captions}
\vskip1pc\noindent
Fig. 1 Energy levels and resonant frequencies of interest.
\vskip0.5pc\noindent
Fig. 2  Entangled state formation, (1): $Re~ D_0$, (2): $Im~D_0$, (3): $Re~D_5$, (4): $Im~D_5$ with
$J'=0.2$
\vskip0.5pc\noindent
Fig. 3 Probabilities for  $J'=0.2$, (k): $|C_k(t)|^2$ for
$k=0,\dots,7$. 
\vskip1pc\noindent
Fig. 4 Expected values (a): $\langle I_0^z\rangle$,  (b): $\langle I_1^z\rangle$, and
(c): $\langle I_2^z\rangle$.
\vskip0.5pc\noindent
Fig. 5 For $J'=0.2$, expected values of the transversal components of the spin.
\vskip0.5pc\noindent
Fig. 6   Real, Imaginary parts of the Fidelity, and its modulus.

\vfil\eject
\leftline{\bf References}
\obeylines{
1. P.W. Shor, {\it Proc. of the 35th Annual Symposium on the Foundation
\quad of the Computer Science}, IEEE, Computer Society Press, N.Y. 1994, 124.
2. L.K. Grover, Phys. Rev. Lett., {\bf 79} (1997) 627.
\quad L.K. Grover, Science, {\bf 280} (1998) 228.
3. P.W. Shor, Phys. Rev. A {\bf 52} (1995) R2493.
\quad E.Knill, R. Laflamme, and W.H. Zurek, Science, {\bf 279} (1998) 342.
A. Stean, Proc. R. Soc. London Se A, {\bf 452} (1996) 2551.
4. F. Schmidt-Kaler, S. Gulde, M. Riebe, T. Deuschle, A. Kreuter, G. Lancaster, 
\quad C. Becher, J. Eschner, H. H\"affner, and R. Blatt, 
\quad J. Phys. B, {\bf 36} (2003) 623.
\quad C.F. Roos, G.P.T. Lancaster, M. Riebe, H. H\"affner, W. H\"ansel, S. Gulde, 
\quad C. Becher, J. Eschner, F. Schmidt-Kaler, R. Blatt, 
\quad Phys. Rev. Lett. {\bf 92},  (2004) 220402.
5. M. Riebe, H. H\"affner, C. F. Roos,Ê W. H\"ansel, J. Benhelm, G. P. T. Lancaster, 
\quad T.W. K\"orber, C. Becher, F. Schmidt-Kaler, D. F. V. James,Ê R. Blatt,
\quad  Nature {\bf 429}, (2004) 734.
6. D. Boshi, S. Branca, F.D. Martini, L. Hardy, and S. popescu
\quad Phys. Rev. Lett., {\bf 80} (1998) 1121.
\quad C.H. Bennett and G. Brassard, {\it Proc. IEEE international Conference on 
\quad Computers, Systems, and Signal Processing}, N.Y. (1984) 175.
\quad I.L. Chuang, N.Gershenfeld, M.G. Kubinec, and D.W. Lung
\quad Proc. R. Soc. London A, {\bf 454} (1998) 447.
\quad I.L. Chuang, N. Gershenfeld, and M.G. Kubinec
\quad Phys. Rev. Lett., {\bf 18} (1998) 3408.
\quad I.L. Chuang, L.M.K. Vandersypen, X.L. Zhou, D.W. Leung, and S. Lloyd,
\quad Nature, {\bf 393} (1998) 143.
\quad P.Domokos, J.M. Raimond, M. Brune, and S. Haroche,
\quad Phys. Rev. Lett., {\bf 52} (1995) 3554.
\quad J.Q. You, Y. Nakamura, F.Nori, Phys. Rev. Lett.,{\bf 91} (2002) 197902.
7. Lloyd, Science, {\bf 261} (1993) 1569.
\quad G.P. Berman, G.D. Doolen, D.D. Holm, and V.I Tsifrinovich
\quad Phys. Lett. A, {\bf 1993} (1994)  444.
8. G.P. Berman, G.D. Doolen, D.I. Kamenev, G.V. L\'opez, and V.I. Tsifrinovich
\quad Phys. Rev. A, {\bf 6106} (2000) 2305.
9. G.P. Berman, G.D. Doolen, G.V. L\'opez, and V.I. Tsifrinovich
\quad quant-ph/9802015, quant-ph/9909032, Phys. Rev. A, {\bf 61} (2000) 062305.
10 . A. Peres, Phys. Rev. A {\bf 30} (1984) 1610.
}

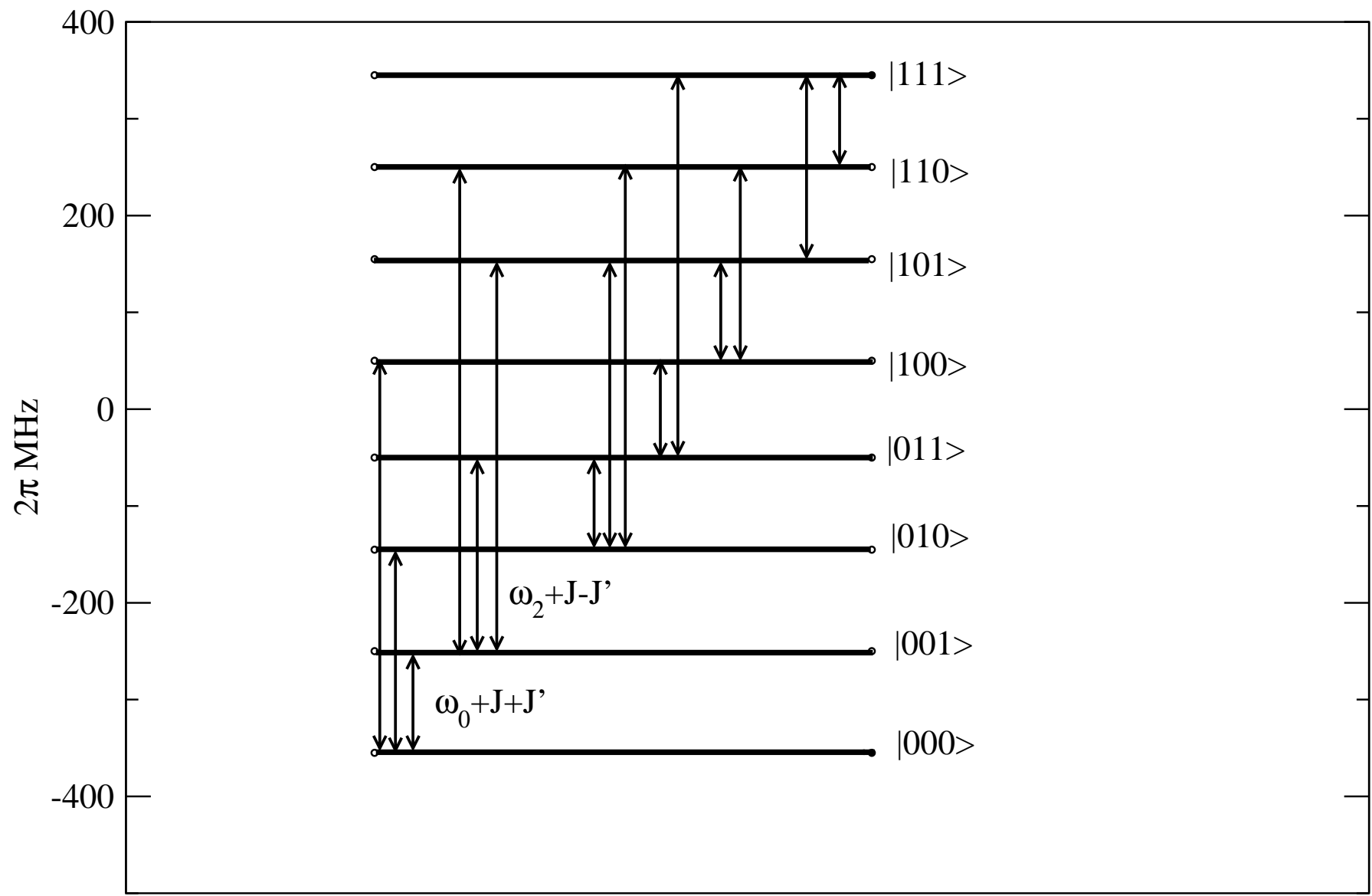
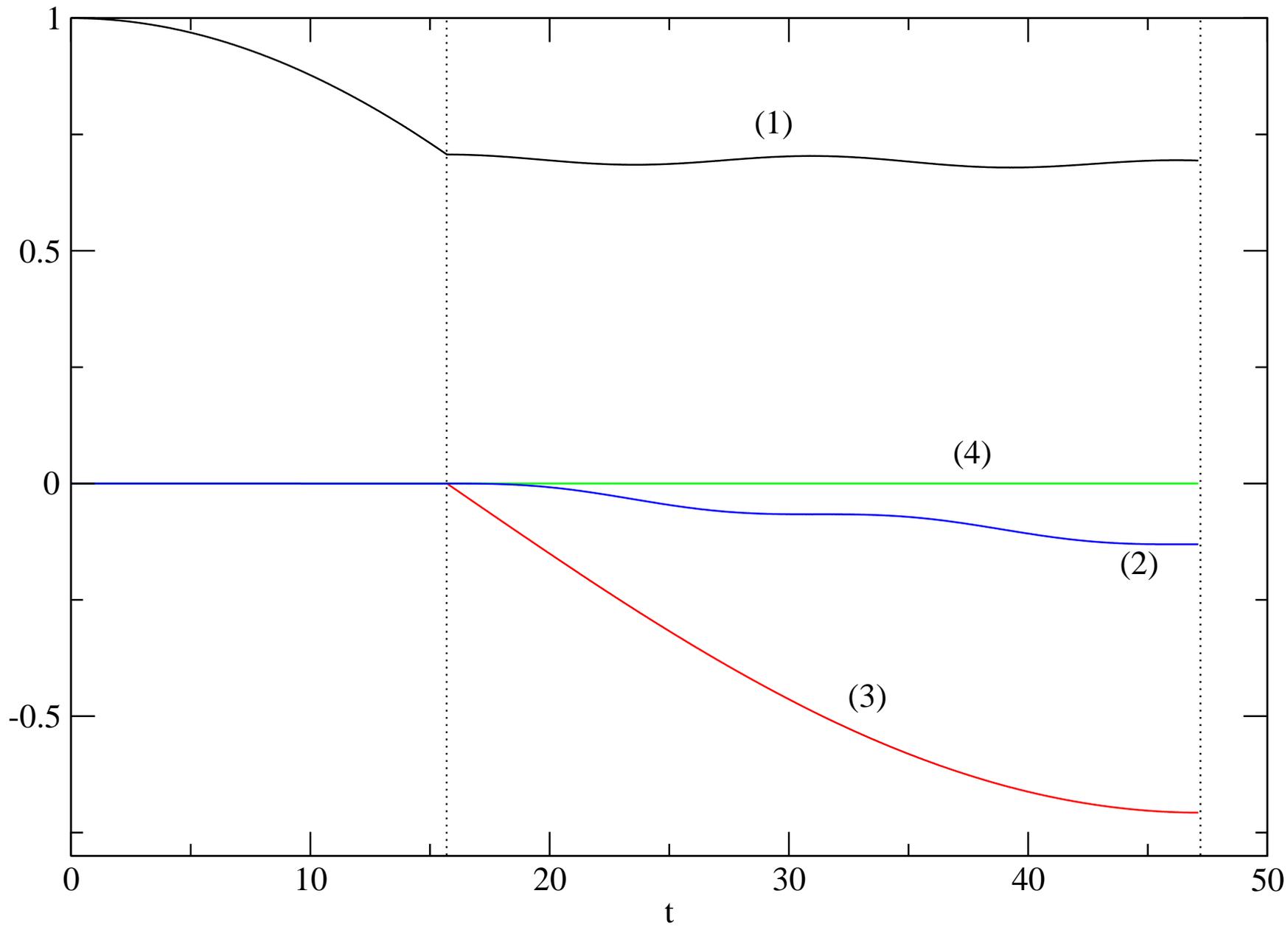
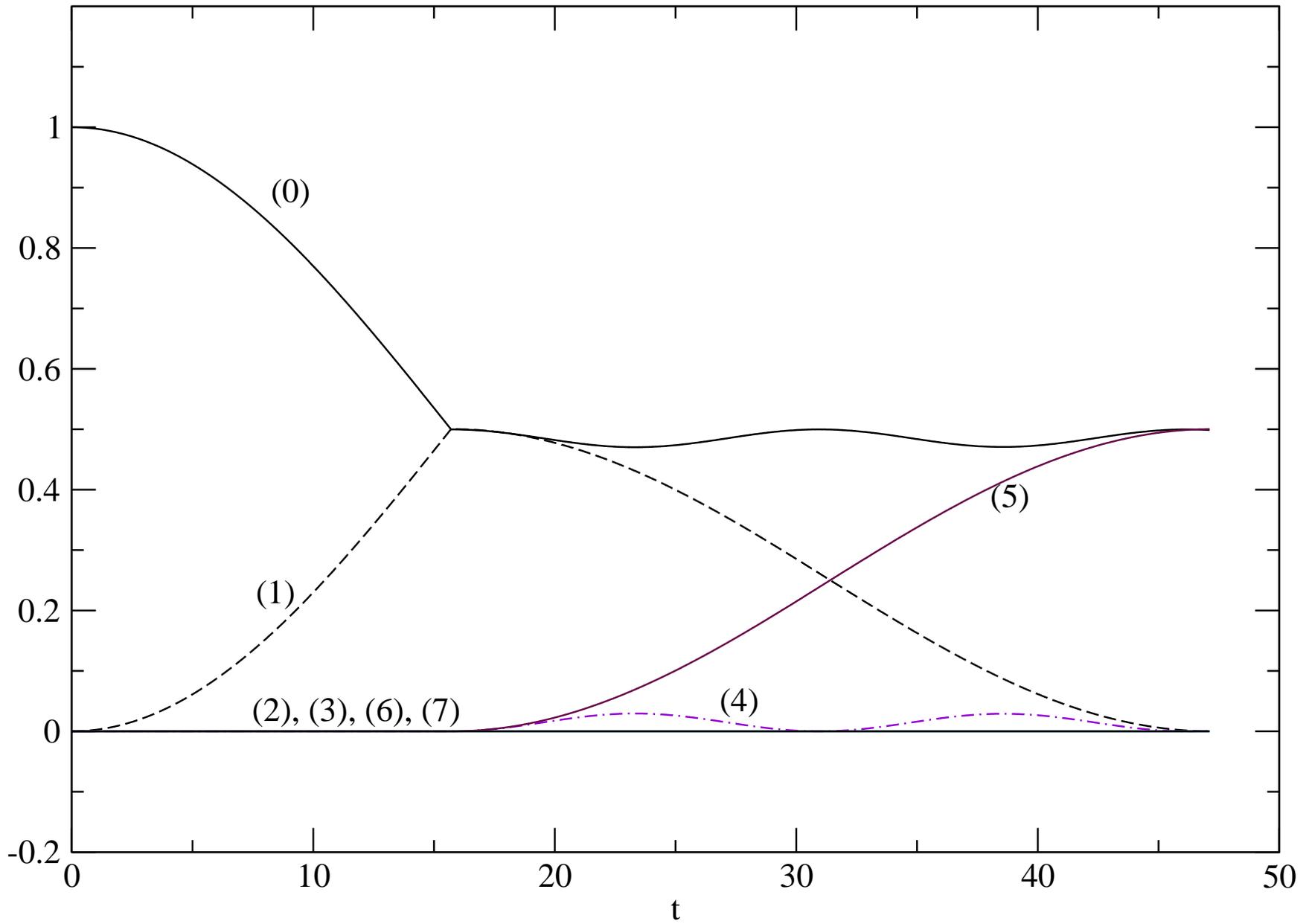
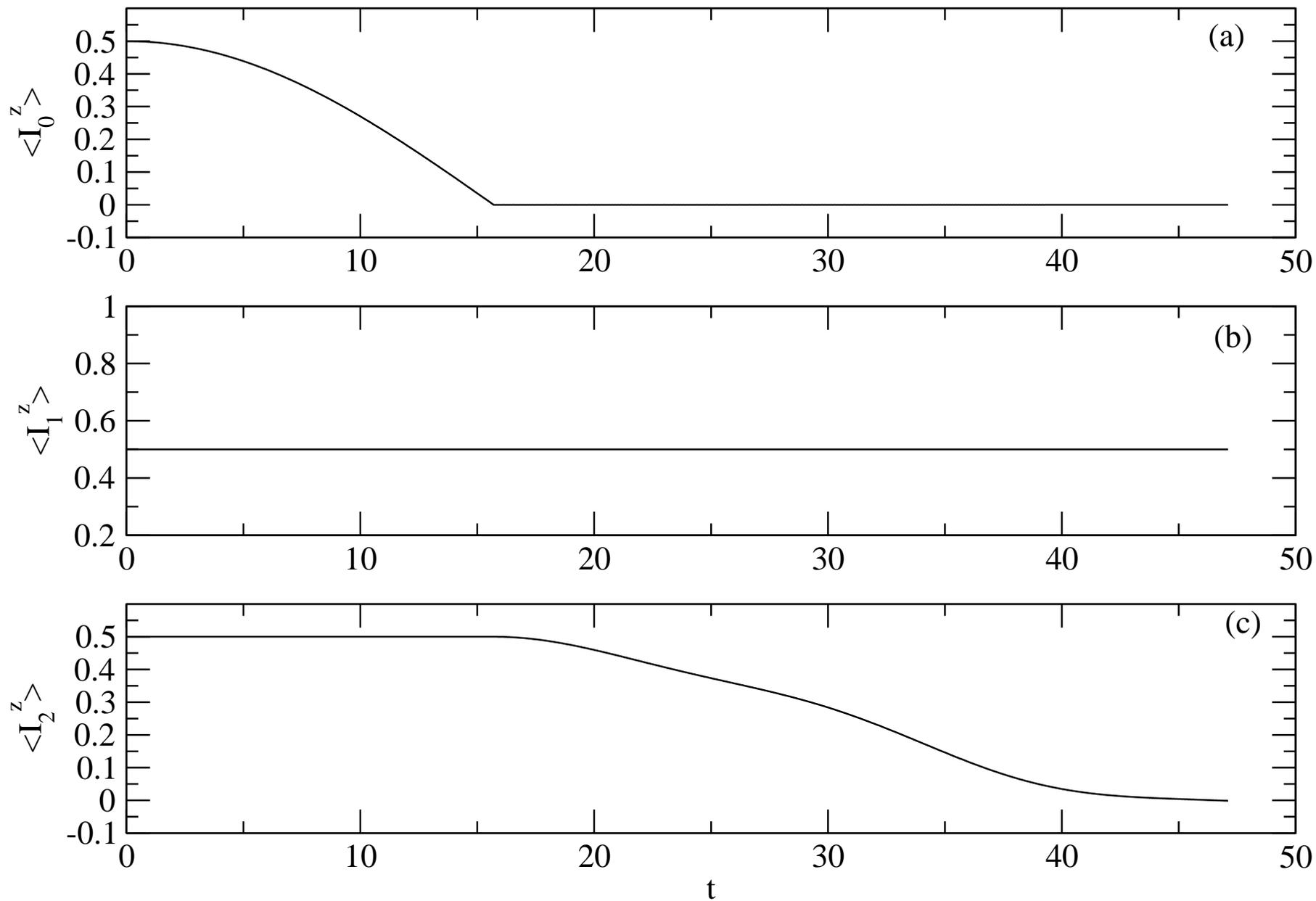
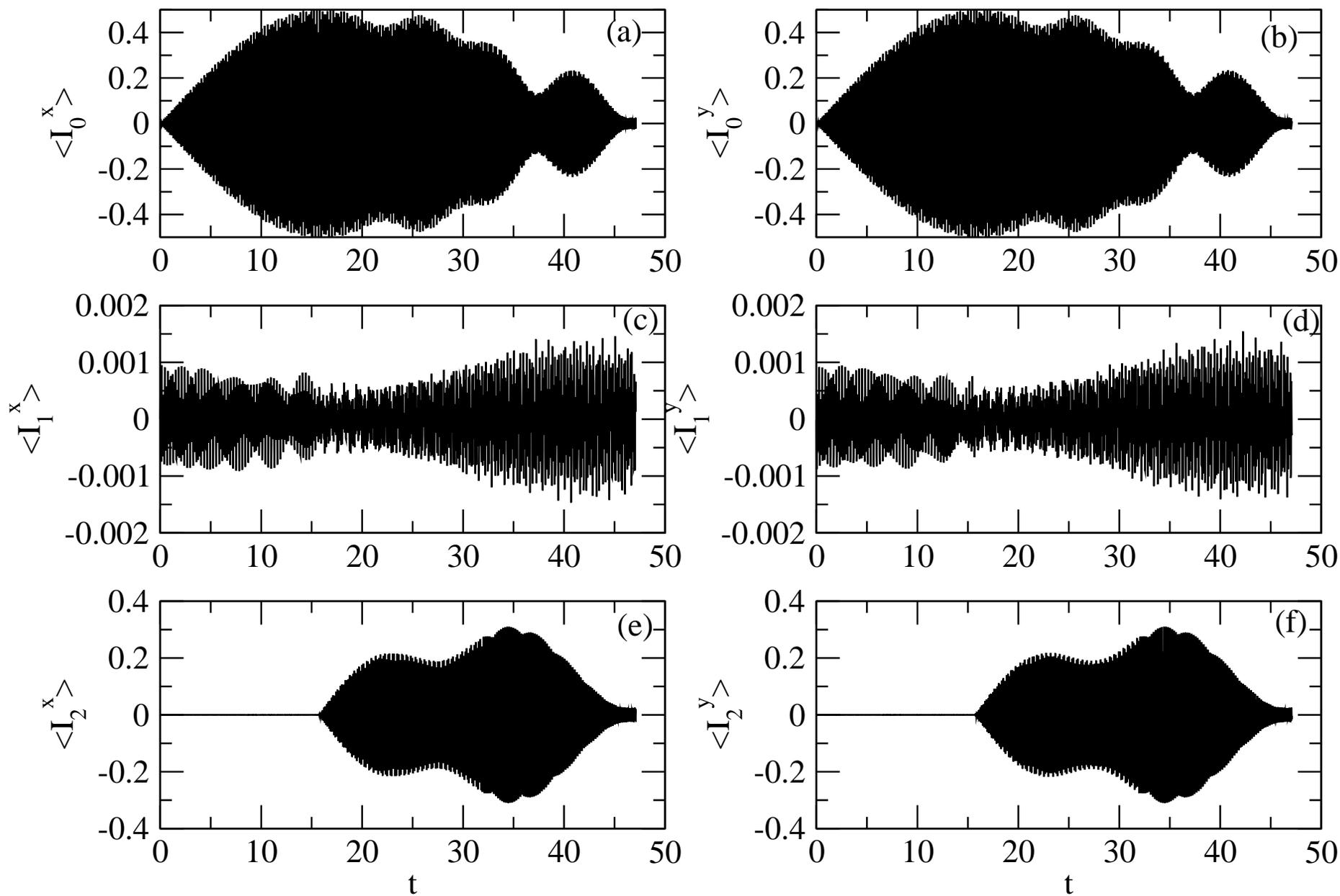
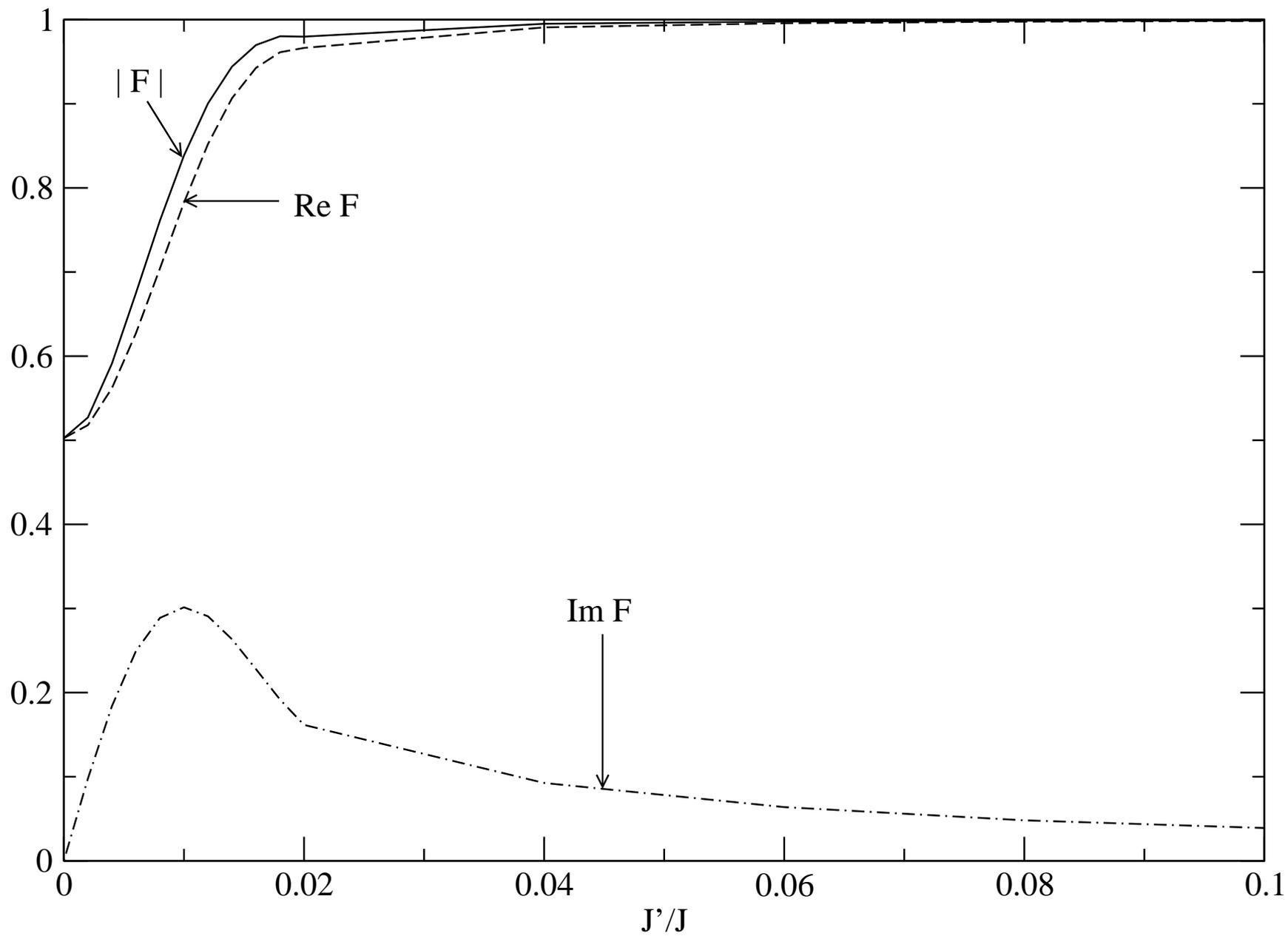
\end{document}